\documentclass[twoside]{dis08}
\usepackage[latin1]{inputenc}
\usepackage[dvips]{graphicx,epsfig,color}
\usepackage{wrapfig,rotating}
\usepackage{amssymb,amsmath,array}

\begin{document}
\title{The dipole picture in DIS: saturation and heavy quarks}

%***********************************************************************
% AUTHORS INFORMATION AREA
%***********************************************************************
\author{Gr\'egory Soyez$^1$
%
% acknowledgement
\thanks{Work done under contract No. DE-AC02-98CH10886 with the US
  Department of Energy.}
%
% DO NOT MODIFY THE FOLLOWING '\vspace' ARGUMENT
\vspace{.3cm}\\
%
% Addresses and institutions
Brookhaven National Laboratory - Physics Department\\
Building 510, Upton, NY 11973 - USA
}
%***********************************************************************
% END OF AUTHORS INFORMATION AREA
%***********************************************************************

\maketitle

\begin{abstract}
  We discuss the description of the proton structure function within
  the dipole factorisation framework. We parametrise the forward
  dipole amplitude to account for saturation as predicted by the
  small-$x$ QCD evolution equations. Contrarily to previous models,
  the saturation scale does not decrease when taking heavy quarks into
  account. We show that the same dipole amplitude also allows to
  reproduce diffractive data and exclusive vector meson production.
\end{abstract}

In these proceedings \cite{url} we shall concentrate on Deep Inelastic
Scattering (DIS) at small $x$. In this regime, the photon-proton
cross-section can be factorised as a convolution between the
wavefunction for a virtual photon to fluctuate into a quark-antiquark
pair and the interaction $T$ between this colourless dipole and the
proton:
\begin{equation}\label{eq:facto}
  \sigma_{L,T}(x,Q^2) = 2 \pi R_p^2 \sum_f \int d^2r\,dz\,
  \left|\Psi_{L,T}(r,z;Q^2)\right|^2\,T(r,x),
\end{equation}
where the factor $2 \pi R_p^2$ arises from integration over the impact
parameter. The photon wavefunction can be computed from perturbative
QED and we are left with the parametrisation of the hadronic dipole
amplitude. To that aim, we usually rely on the observation that the
small-$x$ DIS data satisfy {\em geometric scaling} \cite{kgbs},
meaning that, instead of being a function of both $Q^2$ and $x$, they
appear to be a function of
$\tau=\log(Q^2/Q_s^2(x))=\log(Q^2/Q_0^2)-\lambda\log(1/x)$ only, where
$Q_s(x)$ is known as the {\em saturation scale}. Since $r \sim 1/Q$ in
(\ref{eq:facto}), this property suggests that the dipole amplitude is
a function of $rQ_s(x)$ only.

Since the small-$x$ domain extends down to small $Q^2$, the dipole
amplitude is sensitive to the unitarity bound $T<1$. There are two
broad classes of models which differ by their way to implement that
boundary. The first approach is to use an eikonal form as initially
proposed in \cite{gbw}, followed by more precise analysis to
incorporate DGLAP evolution and masses for the heavy quarks
\cite{gbw+}.

The second approach, that we follow through these proceedings,
is to use predictions directly from the Balitsky-Kovchegov equation
describing the QCD evolution to small $x$. It resums the BFKL 
logarithms of $1/x$ and satisfies unitarity by including saturation
effects.

In contrast with the eikonal models which include it by hand, it has
been proven \cite{geomsc} that the solutions of the BK equation
satisfy the property of geometric scaling. More precisely,
\begin{equation}\label{eq:bkgs}
T(r,x) \overset{rQ_s\lesssim 2}{\propto} \exp\left[-\gamma_c
  z-\frac{z^2}{\kappa\lambda\log(1/x)}\right]
\quad\text{ with }
z = \log\left(\frac{4}{r^2Q_s^2}\right)\text{, }Q_s^2=\left(\frac{x}{x_0}\right)^{-\lambda}\text{ GeV}^2.
\end{equation} 
In this expression, $\gamma_c$, $\lambda$ and $\kappa$ are obtained
directly from the BFKL kernel. The first term in the exponential,
surviving at asymptotically small $x$, satisfies geometric scaling,
while the second term violates geometric scaling and describes how it
is approached when $x$ decreases. Geometric scaling is thus respected
when the second term can be neglected {\em i.e.} in a window extending
up to $z = \sqrt{\kappa\lambda\log(1/x)}$, which extends beyond the
saturation momentum itself. This is an important message one learns
from the BK equation: the saturation effects are not only important
for $rQ_s > 1$ (or $Q^2\le Q_s^2$); they do affect the physical
amplitude at larger values of $Q^2$, in a window growing like
$\sqrt{\log(1/x)}$ and where the dipole amplitude $T$ is significantly
smaller than 1.

\begin{wraptable}{r}{0.65\columnwidth}
\begin{tabular}{|l||c|c|c|c|c|}
\hline
model &  $\gamma_c$  &  $v_c$  &      $x_0$     & $R_p$ & $\chi^2/n$ \\
\hline
\hline
IIM &    0.6275    &  0.253  & 2.67 $10^{-5}$ & 3.250 &$\approx$0.9\\
\hline
IIM+c,b &    0.6275    &  0.195  & 6.42 $10^{-7}$ & 3.654 &
1.109    \\
\hline
new fit\cite{gs}
      & 0.7065 & 0.222 & 1.19 $10^{-5}$ & 3.299 & 0.963 \\
\hline
\end{tabular}
\caption{Values of the parameters and $\chi^2$ per data point for (i)
  the original IIM model, the IIM model with heavy quarks and fixed
  $\gamma_c$ and (iii) the new, adapted, model.}
\label{tab:params}
\end{wraptable}

We thus can use (\ref{eq:bkgs}) to parametrise $T$ at small dipole
sizes and match it continuously with an expression of the form
$1-\exp(-(az+b)^2)$, describing the solutions of the BK equation in
the deep saturation domain. The parameters $\lambda$, $x_0$ and $R_p$
are then fitted\footnote{$\gamma_c$ and $\kappa$ are fixed to the
  value predicted from the leading-order BFKL kernel.} to reproduce
the latest HERA measurements of the inclusive proton structure
function for $x\le 0.01$. This method has been successfully applied by
Iancu, Itakura and Munier (IIM) \cite{iim}, as shown in the first line
of Table \ref{tab:params}, where the sum over quark flavours in
(\ref{eq:facto}) only account for three light quarks.

\begin{wrapfigure}{r}{0.5\columnwidth}
  \centerline{\includegraphics[width=0.49\columnwidth]{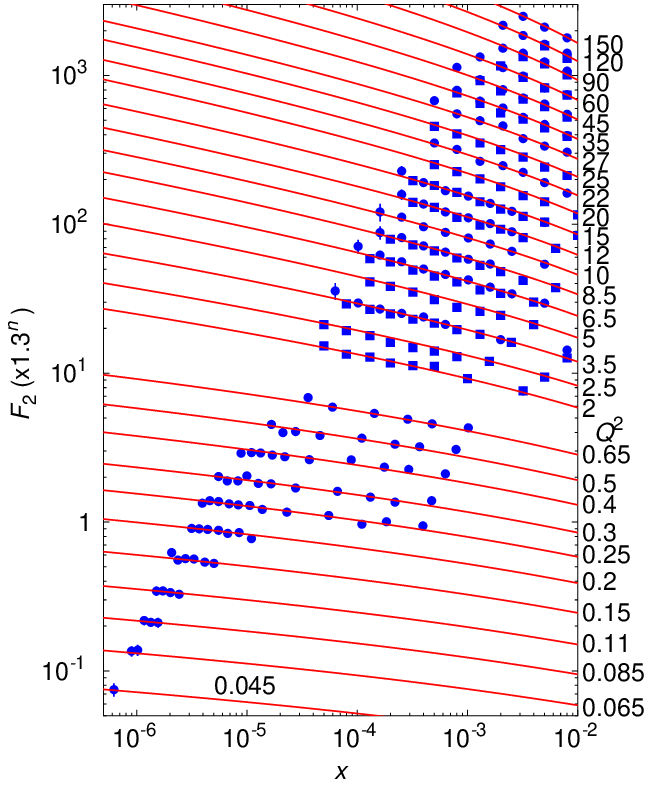}}
  \caption{Description of the $F_2^p$ HERA data.}\label{fig:f2p}
\end{wrapfigure}

One of the general issues of these models is that, in both classes of
models, once the mass of the heavy quarks is taken into account in
(\ref{eq:facto}), the saturation scale drops down by a factor
$\approx$ 2 ($\approx$ 500 MeV instead of $\approx$ 1 GeV). This is
illustrated by the second line of Table \ref{tab:params}, where we see
that, once the contribution from heavy quarks is included, the quality
of the fit becomes poor and $x_0$ decreases severely.

Recently \cite{gs}, I have shown that it was possible to accommodate,
for the first time, the IIM model to include heavy quark contributions
without having the inconvenient that the saturation scale goes
down. The underlying idea is to allow the slope $\gamma_c$ to become a
free parameter of the fit. As shown on the third line of Table
\ref{tab:params}, this does not only brings the parameters closer to
the original IIM model, especially $x_0$ as we will comment further
later on, but it also results in a much better $\chi^2$. To obtain the
parameters mentioned in Table \ref{tab:params}, one has restricted the
$Q^2$ range to $Q^2\le 150$ GeV$^2$, though the parameters are stable
when we vary this limit. Note however that we do expect corrections
from resummation of the DGLAP logarithms at larger $Q^2$.

\begin{wrapfigure}{l}{0.53\columnwidth}
\centerline{\includegraphics[width=0.52\columnwidth]{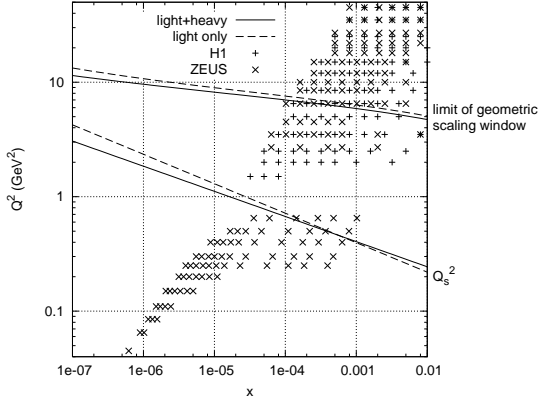}}
\caption{Saturation line and extension of the geometric
  scaling window with HERA data.}\label{fig:qs}
%\caption{Saturation line (bottom lines) and extension of the geometric
%  scaling window (top lines) in the HERA kinematic
%  plane.}\label{fig:qs}
\end{wrapfigure}

Figure \ref{fig:f2p} shows how well the $F_2^p$ data \cite{f2pdata}
are reproduced. More interestingly, we have compared on
Fig. \ref{fig:qs} the saturation scale obtained in this new
parametrisation with the one of the IIM model (bottom lines). They are
clearly of the same order, showing that it is possible to include the
heavy quarks in the dipole picture and at the same time keep a dipole
amplitude with a saturation scale around 1 GeV at HERA. We also
observe on Fig. \ref{fig:qs} (upper lines) that the data up to
$Q^2=5-7$~GeV$^2$ lie inside the geometric-scaling window and are thus
sensitive to the physics of saturation.

\begin{figure}[h!]
\centerline{\includegraphics[width=0.93\columnwidth]{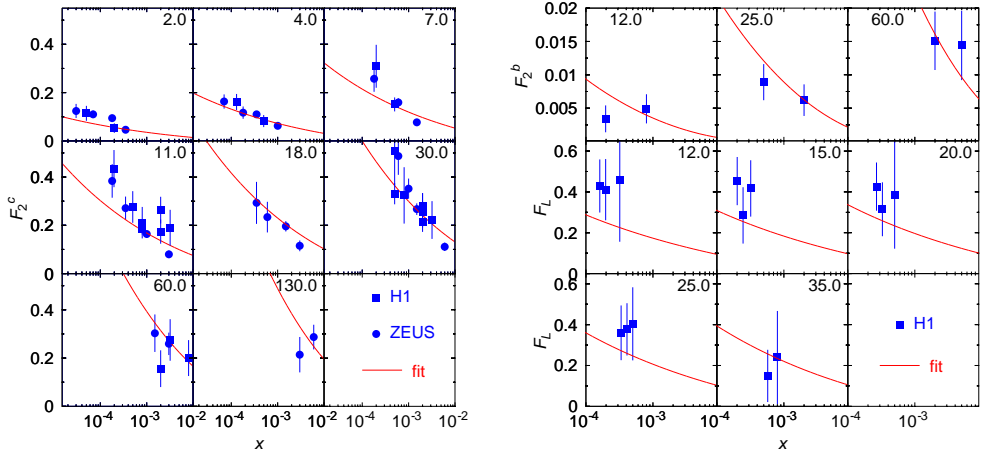}}
\caption{The charm structure function (left plot), bottom structure
  function (top right) and longitudinal structure function (bottom
  right).}\label{fig:othersf}
\end{figure}

Now that heavy quarks are properly integrated into the picture, we can
have predictions for the charm and bottom structure functions. One
sees from Figs. \ref{fig:othersf} that we achieve a good description
of those data. This figure also shows the predictions for the
longitudinal structure function where the model once again agrees with
the data.

With this new model for the forward dipole amplitude that includes
heavy quarks and remains compatible with small-$x$ QCD evolution, we
can also look at exclusive processes. As we show now, one of the major
power of the dipole picture is that it allows, with the same
parametrisation of the dipole amplitude, to describe both inclusive
and diffractive processes.

The first of those measurements we shall consider is the diffractive
structure function. Generally speaking, this quantity probes
correlations inside the proton and is sensitive to the square of the
dipole amplitude. However, considering only fluctuations of the photon
into a colourless $q\bar q$ state only allows to describe the limit of
large $\beta$ (or the limit small diffractive mass $M_X\ll Q$). To go
to smaller values of $\beta$ (keeping $x_{\text{pom}}=x/\beta\ll 1$),
we also have to consider the radiation of one additional gluon from
the initial quark-antiquark pair. We can then show that the
interaction between the resulting $q\bar q g$ pair and the proton can
also be expressed in terms of the dipole amplitude $T$, leading again
to a contribution proportional to $T^2$. Based on the new
parametrisation \cite{gs}, it has been shown \cite{f2d} that the HERA
measurements of the diffractive structure function are well
reproduced.

\begin{wrapfigure}{l}{0.35\columnwidth}
\includegraphics[width=0.35\columnwidth]{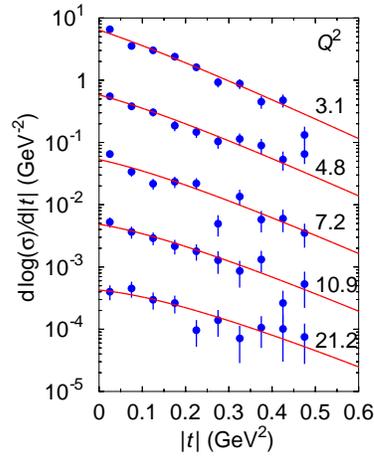}
\caption{Description of the $\rho$-meson
  differential cross-section.}  \label{fig:vm}
\end{wrapfigure}

The final observable we shall consider is the production of exclusive
vector mesons or on-shell photons (DVCS). The differential
cross-section $d\sigma^{\gamma^*p\to Vp}/dt$ can also be factorised,
this time as a convolution between a vertex for the virtual photon to
fluctuate into a $q\bar q$ pair, the interaction between that dipole
and the proton, and the vector-meson wavefunction to account for the
final state. The main difference with (\ref{eq:facto}), beside the
presence of the vector-meson wavefunction, is that one has to account
for the momentum transfer dependence of the dipole scattering
amplitude. An intuitive method is to make a Fourier transform and go
to impact-parameter space. We shall rather use the result of studies
of the full BK equation including its momentum-transfer dependence. It
is predicted \cite{bkfull} that the saturation scale is 
constant at small $t$ and increases like $|t|$ at large
$t$. Introducing one parameter to implement that dependence and a
second one to describe the proton form factor (taken of the form
$\exp(-b|t|)$), we have reached \cite{vm} a successful description of
the differential and total cross-sections for exclusive productions of
$\rho$, $\phi$ and $J/\Psi$ mesons, as well as the for DVCS
measurements. The description of the $\rho$-meson differential
cross-section is given as an example on Fig. \ref{fig:vm}.

% ****************************************************************************
% BIBLIOGRAPHY AREA
% ****************************************************************************

\begin{footnotesize}
% IF YOU DO NOT USE BIBTEX, USE THE FOLLOWING SAMPLE SCHEME FOR THE REFERENCES
% ----------------------------------------------------------------------------

% ----------------------------------------------------------------------------

\end{footnotesize}

% ****************************************************************************
% END OF BIBLIOGRAPHY AREA
% ****************************************************************************

\end{document}